\documentclass{article}
\usepackage[utf8]{inputenc}
\usepackage{graphicx}

\title{\textbf{Sleep syndromes onset detection based on automatic sleep staging algorithm}}
\author{Tim Cvetko, Tinkara Robek }
\date{May 2021}

\begin{document}

\maketitle

\section{Abstract
}
Sleep specialists often conduct manual sleep stage scoring by visually inspecting the patient’s neurophysiological signals collected at sleep labs. This is, generally, a difﬁcult, tedious and time-consuming task. The limitations of manual sleep stage scoring have escalated the demand for developing Automatic Sleep Stage Classiﬁcation (ASSC) systems. Sleep stage classiﬁcation refers to identifying the various stages of sleep and is a critical step in an effort to assist physicians in the diagnosis and treatment of related sleep disorders.
In this paper, we propose a novel method and a practical approach to predicting early onsets of sleep syndroms, including restless leg syndrome, insomnia, based on an algorithm which is comprised of two modules. A Fast Fourier Transform is applied to 30 seconds long epochs of EEG recordings to provide localized time-frequency information, and a deep convolutional LSTM neural network is trained for sleep stage classification. Automating sleep stages detection from EEG data offers a great potential to tackling sleep irregularities on a daily basis. Thereby, a novel approach for sleep stage classification is proposed which combines the best of signal processing and statistics. In this study, we used the PhysioNet Sleep European Data Format (EDF) Database. The code evaluation showed impressive results, reaching accuracy of 90.43, precision of 77.76, recall of 93,32, F1-score of 89.12 with the final mean false error loss 0.09. All the source code is availlable at\textit{ https://github.com/timothy102/eeg. }

\subsection{Index Terms: }
Sleep stage classification, algorithm design, onset detection, convolutional neural networks, signal processing, recurrent neural networks, fourier transform, CEEMDAN, preprocessing

\section{Introduction
}
Sleep stage classification is crucial for our day-to-day functioning and disease prevention, such as obesity, type 2 diabetes, and different psychiatric disorders that may develop in association with the shortage of sleep.Maintaining a healthy sleep schedule is significant for obtaining our natural circadian rhythm and the normal course of sleep. This must include REM (rapid-eye movement) and non-REM phases.  NREM is specified by slow-wave activity (delta power). REM phase is characterized by the eye movements which we track with EOG (electrooculogram), low muscle tone which we track with EMG(electromyogram), low amplitude waves, and generally mixed frequency EEG (electroencephalogram). It is also a stage where dreams occur. The brain is as active during REM as during wakefulness, only in different parts (correspondingly, the prefrontal cortex, responsible for judgment and decision making is deactivated during both phases). The REM  state is initialized by the neurons located in the brainstem.  The process of recording sleep in the laboratories for the purposes of examining its structure (mainly in terms of slow-wave activity) is known as polysomnography. Polysomnograms can show the diagnostics of various disorders during sleep. They are commonly used for identifying disorders such as obstructive sleep apnea, excessive daytime sleepiness, and insomnia. Polysomnograms record numerous types of information: EEG (electroencephalogram; for recording brain waves), EOG (electrooculogram; for recording eye movements), EMG (electromyogram; recording of the electrical activity of muscle surfaces), EKG (electrocardiogram). Combining the gathered data allows us to identify the individual's stage of sleep. For the purposes of recognizing sleep apnea, it is also required to record snoring, nasal and oral airflow, movement of the abdomen and chest, and pulse oximetry. There are multiple types of waves measured in EEG which tell whether the person is asleepor awake and what stage they are in. Alpha activity (frequency: 8-13Hz) is present in the occipital cortex when a person is resting with their eyes closed and goes away when they wake up or fall asleep. Theta activity (4-7Hz) is the most common frequency during sleep. Delta activity is known as slow-wave activity (0,5-2/4Hz) with higher amplitudes and is mostly present in the frontal areas of the cortex. Sleep spindles (11-16Hz) normally indicate stage 2 of NREM sleep, with a duration of 2-3 seconds. K complexes are also most common in stage 2 of NREM, while they might appear in other phases as well. Their amplitude is not specified, while the time span must be at least 0.5 seconds.Sleep stages are regulated by individual neurotransmitters. Wakefulness is a heterogeneous state, maintained by neurotransmitters acetylcholine, serotonin, norepinephrine, histamine, dopamine, hypocretin, and glutamate. NREM is initialized by GABA and adenosine, while acetylcholine also sends signals for REMonset. Interactions between neurotransmitters determine behavioral states at any given time. During sleep, they define muscle tone, eye movements and EEG activity. The same neurotransmitter can have opposite effects during wakefulness and sleep and in different brain regions. The preoptic area of the brain (anterior hypothalamus) is known to act as a sleep center, while the posterior hypothalamus acts as a wakefulness center. Electrical stimulation of the preoptic area, therefore, has the ability to initiate sleep.  Apart from the objective reasons for not sleeping long enough  (the recommended 8 hours, for obtaining full sleep cycles), there are also multiple sleeping disorders present in today's society. Sleeping disorder is usually a composition of multiple neurological modifications that may be caused by external or internal factors. These factors include our lifestyle, genetics, associated diseases, psychiatric disorders, and so on. Disordered sleeping is alarmingly common in modern society. . In the field of various disorders, the most common persist to stay the following: insomnia, sleep-disordered breathing (obstructive sleep apnea, hyperventilation), central disorders of hypersomnolence (narcolepsy, idiopathic hypersomnia, Kleine-Levin Syndrome), circadian rhythm disorders, parasomnias (sleepwalking, terrors, sleep eating disorders, sleep enuresis) and sleep-related movement disorders (restless leg syndrome, periodic limb movement disorder). Insomnia and other disorders that result in trouble sleeping are often associated with multiple chronic health conditions. They have an impact on blood pressure and blood sugar levels, therefore they put us at risk of coronary diseases and metabolic imbalances. Middle-aged/older parts of the population and obese individuals are at grater risk of developing these issues. In addition, sex also plays a role in developing some sleeping disorders - males are more likely to develop obstructive sleep apnea, while women are predisposed to insomnia. The first step for identifying problems with sleep is by clinical approach with ESS (EpworthSleepiness Scale)[0]. Insomnia can arise as chronic insomnia or in a short-term form. Its typical symptoms are problems initiating or maintaining sleep, non-restorative sleep, and waking too early which result in daily distress. These issues occur due to the brain's arousal systems which are not turning off.  It is diagnosed in a laboratory with polysomnography. It can be a cause for endothelial dysfunction, oxidative stress, inflammation, hypertension, and even stroke in severe cases. Ordinarily, it is treated with CPAP (continuous positive airway pressure).

\section{Methodology
}

\subsection{Event detection during sleep}

Although detecting periods of raised or lowered activity during sleep is an appealing task, it is rather interesting to detect transient, short signals. Furthermore, during sleep we can for example detect events like QRS complexes in the ECG or EEG events like slow waves (single waves of about 0.5 to 2 Hz which have a high amplitude of about 75 uV), sleep spindles (transient waxing and waning events of about 0.5 to 2 seconds duration and 15 to 50 uV maximal amplitude). For the sake of this research and simplicity we do not make any distinction between slow-waves, K-complexes or slow oscillations but lump them all together. 
\newline
\subsection{Detailed description of the methodology}

For the preprocessing part, we took advantage of the complete ensemble empirical mode decomposition with adaptive noise (CEEMDAN) as a noise assisted EMD  technique. Several signal processing methods have been developed in recent years. The traditional signal processing method is based on the Fast Fourier Transform. However, this method alone cannot obtain the time-domain and frequency-domain analyzed results simultaneously. Many time-frequency analysis methods have been applied to diagnoses for vibration signals, such as the Wigner-Ville distribution (WWD) as proposed in [2]. This algorithm successively computes the analytic signal corresponding to the input signal, forms a weighted kernel function, and analyzes the kernel via a discrete Fourier transform (DFT). To evaluate the analytic signal required by the algorithm, it is shown that the time domain definition implemented as a finite impulse response (FIR) filter is practical and more efficient than the frequency domain definition of the analytic signal, according to [3]. The wavelet transform ([4], [5]) utilizes the Fast Fourier  Transform to extract frequency based wavelets. Empirical mode composition is one of the most powerful signal processing techniques used for fault diagnosis. Deriving from local characteristic timescales of a signal and can decompose the signal into a set of complete and almost orthogonal components known as intrinsic mode function (IMF). The IMFs indicate the natural oscillatory mode imbedded in the signal and serve as the basis functions, which are determined by the signal itself, rather than by predetermined kernels. Therefore, it is a self-adaptive signal processing technique that is suitable for nonlinear and nonstationary processes and is suitable for fault feature extraction of spiral-bevel gears. According to [5], CEEMDAN has been proven to be an effective method for analyzing nonstationary signal followed by strong noise. First, the raw vibration signal is decomposed by the CEEMDAN method to obtain a series of intrinsic modal functions (IMFs). The IMFs which included greater amounts fault information are selected as the optimal IMFs based on the correlation coefficient. Next, the permutation entropy values of the optimal IMFs are calculated. In order to obtain accurate permutation entropy values, the two key parameters, namely, embedding dimension and delay time, are optimized by using the overlapping parameter method. In order to assess the sensibility of the permutation entropy features, the support vector machine (SVM) is used as the classifier for fault mode identification, and the diagnostic accuracy can verify its sensibility.

\hfill \break
 \subsection{ Model architecture}
   The standard practice for image recognition, computer vision and more are deep convolutional neural networks. Models like, LeNet[7], AlexNet[8], ImageNet[9] and more work exclusively on 2D data. They are also referred to as 2D CNNs. With the same idea in mind, 1D CNN have been developed. Despite the obvious fact, that the 1D CNNs apply a 1D kernel to data, some of the advantages are:
1D CNNs don’t require heavy matrix operations ( nor at the forward pass or the backpropagation). Potentially, what this means is that they are, compared to their 2D counterparts, computionally less complex and can run on CPU hardware. 

In this study, we have come up with a twin fully connected convolutional network. Each convolutional block consists of a convolutional 1D layer, which is passed through a Leaky Rectified Linear Unit as a nonlinearity. The input signal then passes a max-pooling layer with the stride of 2, following that is batch normalization layer and finally the Bidirectional layer. Each twin structure contains three such blocks. The outputs of the convolutional-LSTM blocks are then concatenated across axis 1 in order to be finally passed through a fully connected 64 unit dense layer. We use the softmax activation to recieve the final scores for each sleep stage. The CNN consists of three different-sized filters with the idea in mind being that small filters extract temporal information and large filters to extract frequency information. The idea behind these variable sizes of filters comes from the signal processing community to have a trade-off between extracting time domain
and frequency domain features [6]. This helps get benefit from both time and frequency domain features in the classification task. The feature map generated from the 1D concatenated output from the twin convolutional blocks are intertwined with long short-term memory (LSTM) units to capture the complex context dependencies between the inputs and the targets.

\subsection{Utilizing the Conv-1D and the Bidirectional LSTM}
The Conv-LSTM network combines the best of both worlds, one-dimensional convolutional layers handles spatial data using a kernel to train a set of filters that best represent the hyperplane and the LSTM which encodes temporal inputs using cell state memory. To put it differently, Conv-LSTM predicts the future state of a cell given vectors on a spatial grid and past states of neighbours.

Convolutional-LSTM layer works by passing the 1D kernel over a set of training points and its result a.k.a feature map is passed onto the LSTM cell. In this project, we have utilized the bidirectional recurrent neural network instead of the standard LSTM cell. Put simply, this was enabled due to the limitations of LSTM having the limitation of being restricted from processing previous input state. Given the inputs x = {x0, xN} , which are single dimensional data of a 30-second clip, which should assure stability and little overlap, these inputs are passed into the twin architecture.

\begin{figure}
  \centering
 \includegraphics[width=\columnwidth]{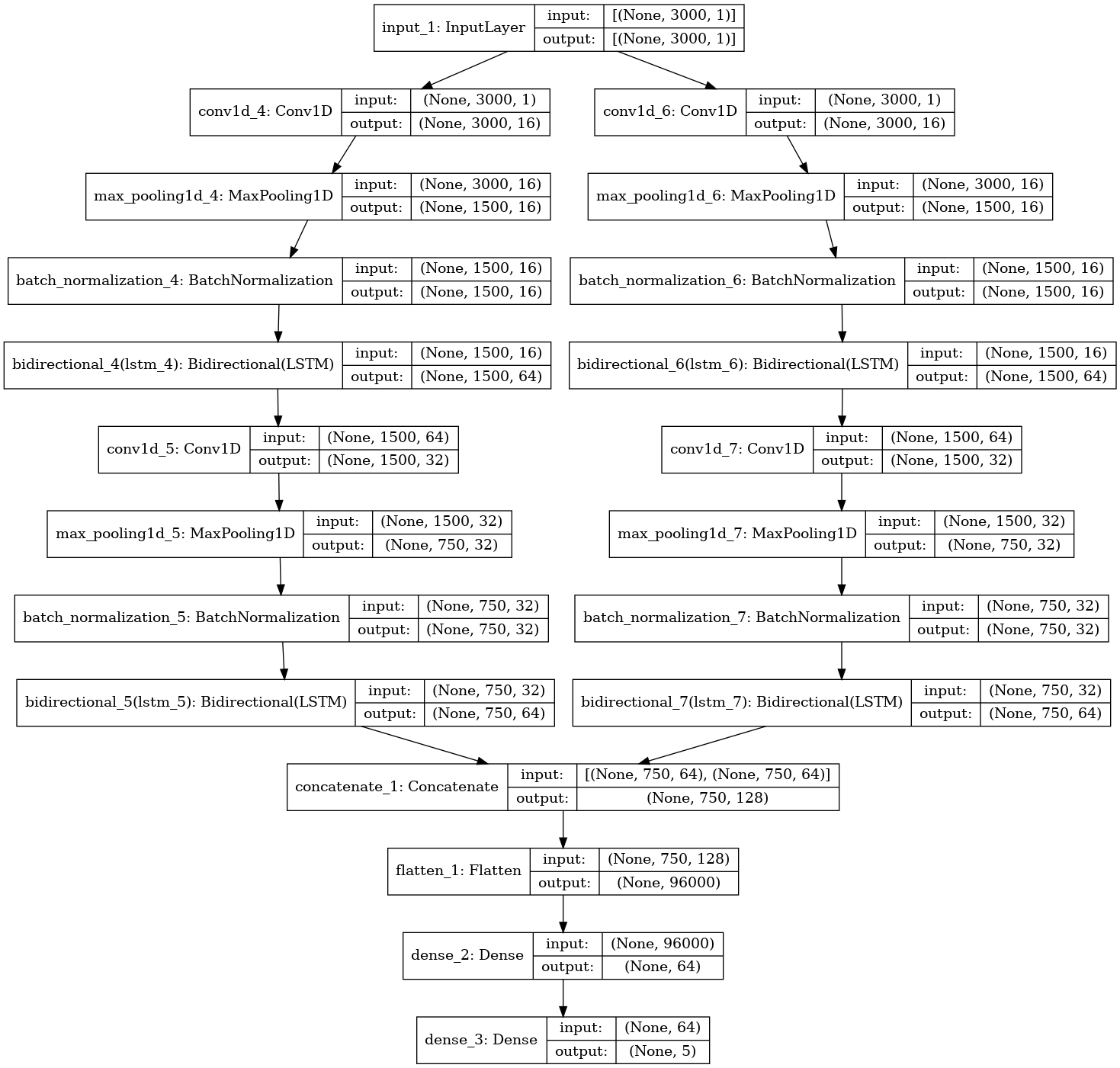}
  \caption{A schema of the twin convolutional Bidirectional LSTM network}
  \label{fig:model}
\end{figure}
\subsection{Loss calculation}

Quite obviously, sleep stage classification encounters the class imbalance issue. To alleviate this issue, we have taken both the extended mean false error(MFE) and the mean squared false error(MSFE) for multiclass classification.

\hfill \break
$MFE_L = \sqrt{ \frac{1}{N}\sum\_{i=1}^{C} (x_{i}-\hat{x})^2}$

\hfill \break
\subsection{Unsupervised method for detecting onsets of sleep syndromes}
Syndromes in question are restless leg syndrome(RLS), rapid Eye Movement (REM) sleep behaviour disorder, insomnia, sleep apnea and narcolepsy. In attempt to detect these complex syndromes through EEG data, we have come up with the following: 
\subsubsection{Restless leg syndrome}
A dopaminergic drug - lisuride exhibited increase in alpha, decrease in beta and slow activities on brain function measured by computerized EEG. It was postulated that reverse EEG changes might play role in the pathogenesis of RLS. During transition from wakefulness to sleep stage 1 changes in alpha activity initiate long-lasting alpha arousal responses and they continue increasingly at sleep stage 2. This dysfunction is probably due to a genetic vulnerability of EEG alpha rhythm and disinhibits the diencephalospinal dopamine system, mostly during sleep but also during wakefulness. The disinhibition produces background for activation of PLMs, disturbing sensations in brainstem and urge to move, motor restlessness at cerebral cortex, generally for legs. All lead to severe insomnia. In RLS patients, forced deviations from alpha to theta or beta activity are unsuitable and resting EEGs reflect a dopamine receptor-specific 'individual sensitivity.' This vulnerability is alleviated after lisuride with suitable CEEG changes.

We have designed an unsupervised statistical inference model for approximating the restless leg syndrome in EEG. First the Savitzky–Golay filter was applied to each 30-second window for the purpose of smoothing the data, that is, to increase the precision of the data without distorting the signal tendency. This is achieved, in a process known as convolution, by fitting successive sub-sets of adjacent data points with a low-degree polynomial by the method of linear least squares. Next, an orthnogonal discret Fourier Transform was applied to the data to decompose functions depending on space or time into functions depending on spatial or temporal frequency. At last, the final RLS score is computed by applying the sigmoid activation function to the fourier-transformed data. 

\subsubsection{Sleep apnea}

When a person is sleeping, due to pause in breathing, carbon dioxide may build up in the bloodstream. When this reaches a critical level, it is detected by chemoreceptors. These receptors signal an alarm to the brain to wake the sleeping person and breathe in air. As a result, a transition in sleep stages occurs, which in turn, causes fluctuation in activity level of various frequency bands of the EEG signal. Hence, more distinct features can be preserved in frequency band-limited signals for apnea detection compared with that in full-band EEG signal. As a result, for apnea event detection, characteristics of band-limited signals are used instead of analysing full-band EEG signal. EEG signal is partitioned into five frequency bands [10] including delta (0.25–4 Hz), theta (4–8 Hz), alpha (8–12 Hz), sigma  (12–16 Hz) and beta (16–40 Hz). Spectral filtering is done in fast Fourier transform domain to achieve this division.

In order to attain a relative score for approximating sleep apnea possibility, we have adopted the method proposed in [7] .For classification of apnea and non-apnea events, a subject-specific classification scheme based on feature is introduced in this proposed method. Inter-band energy ratio of band-limited EEG signal, expected to possess differentiating characteristic for apnea and non-apnea events, is proposed as the feature. During sleep apnea, the energy contents in various frequency bands change significantly with respect to non-apnea events.

\subsubsection{REM behaviour sleep disorder}
Despite what was presented in [9], we have come about a rather simple, but clever way of detecting the REM instability score. A relative score was calculated by using a greedy algorithm for longest continuing sequence [10] to obtain a value that should correspond to the disturbance of REM sleep in an individual. 

\subsubsection{Narcolepsy}
According to [11] patients with narcolepsy showed lower alpha power, higher delta and theta power during wakefulness, and higher alpha and beta power during rapid eye movement (REM) sleep. The former two groups also showed lower sleep efficiency and a higher rate of positivity of REM-related symptoms than the other two groups.
With this idea in mind, we have combined the REM instability score and the obtained frequency range with wavelet transform on the first and last 10 percent of sleep, indicating worse sleep-to-wake transitions or vice versa. 
\subsubsection{Insomnia}
Insomnia or sleeplessness is a sleep disorder in which there is an inability to fall asleep or to stay asleep as long as desired. Following the previous work done by Mohd Maroof Siddiqui,  Geetika Srivastava, and Syed Hasan Saeed in [11]. Now each clipped signal is preprocessed and then passed through the Hanning window low pass filter for removing the high frequency components that eventually indicate noise because major proportion of EEG signals are limited within the range of 25 Hz. Hence, the filter based in FIR filter design of order 200 with cut off frequency of 25 Hz with shape of hanning window. At this stage, we adopt a  method for extracting frequency length windows via the Welch Power Spectrum [12]. This score, obtained with softmax and the REM instability score combine for the insomnia score. 

\subsubsection{Sleep depth}
Given a great sleep stage classifier, we concluded to not only move in the direction of detecting abnormalities, but to also quantify sleep depth. First, the raw records are pushed through a Savgol filter to noise filtering. Furthermore, a Fast Fourier Transform has given us a distribution of frequencies. At this stage, we evaluted the FFT distribution to obtain the sleep depth score along with the REM instability score. 

\begin{figure}
  \centering
 \includegraphics[width=\columnwidth]{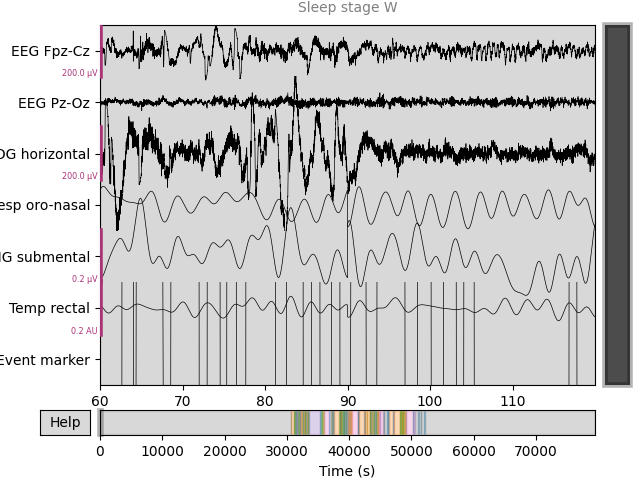}
  \caption{Polysomnography plot for a single patient. Picture was generated with the MNE Python library}
  \label{fig:model}
\end{figure}

\begin{center}
 \begin{tabular}{||c c c c c c c ||} 
 \hline
 Syndromes & RLS & REM disorder & Apnea & Narcolepsy  & Insomnia & Sleep depth \\ [0.5ex] 
 \hline\hline
  In total & 177983 & 6674 &8899  &4449 & 48945 & 164738 \\  
 \hline
 Percentage & 7.80 & 0.29 & 0.41 & 3,0 & 22 & 74.0 \\  
 \hline
 Percentage given NCHS & 7-10 & 0.5-1 & 0.5 & 3-7 & 10-30 & 66 \\
 \hline\hline
\end{tabular}
\end{center}

\section{Experimental results
}

\subsection{Data}
For the purposes of this research, we used the Physionet Sleep-EDF dataset. The sleep-edf database contains 197 whole-night PolySomnoGraphic sleep recordings, containing EEG, EOG, chin EMG, and event markers. Some records also contain respiration and body temperature.The Sleep-EDF dataset contains two different studies including study of age effects on sleep in healthy individuals (SC = Sleep Cassette) and  study of temazepam effects on sleep (ST = Sleep Telemetry). The dataset includes whole-night polysomnograms (PSGs) sleep recordings at
the sampling rate of 100 Hz. Each record contains EEG (from Fpz-Cz and Pz-Oz electrode locations), EOG,
chin electromyography (EMG), and event markers. Few records often also contain oro-nasal respiration and rectal
body temperature. The hypnograms (sleep stages; 30-s epochs) were manually labeled by well-trained technicians
according to the Rechtschaffen and Kales standard. Each stage was considered to belong to a different class
(stage). The classes include W, REM, N1, N2, N3, N4, M. Data syndrome comparison was done according to [13]

\begin{center}
 \begin{tabular}{||c c c c c c c ||} 
 \hline
 Dataset & W & N1 & N2  & N3-N4 & REM & Total \\ [0.5ex] 
 \hline\hline
  Sleep-EDF-13 & 8285 & 2804 & 17799 & 5703 & 7717 & 42308 \\  
 \hline
 Sleep-EDF-18 & 65951 & 21522 & 96132 & 13039 & 25835 & 222479 \\  
 \hline\hline
\end{tabular}
\end{center}

It is worth noting that the sleep stages in the Physionet's Sleep-EDF database are not normally distributed. There is a significantly higher number of W and N2 stages than others which is causing class imbalance. We have discussed this at Loss calculation. 

The model was trained for 20 epochs. We used the Adam optimizer to minimize the MFE loss with a batch size of 16 and a learning rate alpha = 0.001. We also added the L2 regularization penalty with beta being = 0.001 to mitigate the overfitting. For this goal, we used Python as the programming language and TensorFlow as the framework. 

By employing the Conv-1D and the Bidirectional LSTM twin network, we have succesfully taken advantage of the temporal and spatial component of the signal processing nature. To reduce the class imbalances problem, we adopted the weighted MFE loss. 

\begin{center}
 \begin{tabular}{||c c c c c||} 
 \hline
 Accuracy & Precision & Recall  & F1-score & Final MFE \\ [0.2ex] 
 \hline\hline
 90.43 & 77.76 & 93.32 & 89.12 & 0.09 \\  
 \hline
\end{tabular}
\end{center}

.
\section{Conclusion and Future Work
}
In this paper, we have successfully proposed a new deep learning approach to the challenging problem of sleep stage classification that is essential for further analysis of the human brain activity. We utilized a two-part convolutional-LSTM block along with a multi-layer perceptron to classify sleep stages from EEG data with the accuracy of . By incoorporating a vastly important CEEMDAN preprocessing method, we build a more stable and noise-free variant of our data, which enabled building an end-to-end trainable model for not only classifying sleep stages, but also detect early onsets of frequently occuring sleep syndroms. Our algorithm takes advantage of both the spatial and temporal dimension of signal processing, as well as statistical analysis to come up with a score that represents the sleep stage the most. As a limitation, we would like to add that no data is known to bare diluted patients. For future work, given proper data, we would like to do a follow-up. In the future, we intend to extent this using multimodal polysomnography to enhance the model’s performance and usability. Following our work, we will investigate how these syndrome predictions/scores align with actual scores and will hopefully serve as a professional service for automatic sleep stage and syndrome onset detection. 
    
\section{ References
}

    [1] https://epworthsleepinessscale.com/about-the-ess/

[2] Shie Qian and Dapang Chen, "Decomposition of the Wigner-Ville distribution and time-frequency distribution series," in IEEE Transactions on Signal Processing, vol. 42, no. 10, pp. 2836-2842, Oct. 1994, doi: 10.1109/78.324750.

[3] B. Boashash and P. Black, "An efficient real-time implementation of the Wigner-Ville distribution," in IEEE Transactions on Acoustics, Speech, and Signal Processing, vol. 35, no. 11, pp. 1611-1618, November 1987, doi: 10.1109/TASSP.1987.1165070.

[4] Katsuhiro Inoue, Akihiko Takajo, Makoto Maeda and Shigeaki Matsuoka, "Tuning method of modified wavelet transform in human sleep EEG analysis," 2007 International Conference on Control, Automation and Systems, 2007, pp. 2784-2789, doi: 10.1109/ICCAS.2007.4406842.

[5] Xujing Yao, Ji Han, COVID-19 Detection via Wavelet Entropy and Biogeography-Based Optimization, COVID-19: Prediction, Decision-Making, and its Impacts, 10.1007/978-981-15-9682-7-8, (69-76), (2021).

[6] https://www.hindawi.com/journals/sv/2019/7806015/.
[7] Deep Convolutional Neural Networks for Image Classification: A Comprehensive Review - Scientific Figure on ResearchGate. 

Available from: https://www.researchgate.net/figure/Architecture-of-LeNet-5-LeCun-et-al-1998-fig2-317496930 [accessed 3 May, 2021]

[8]  M. X. Cohen, Analyzing neural time series data: theory and practice. MIT press, 2014.

[9] Krizhevsky, A., Sutskever, I. and Hinton, G. E. (2012). ImageNet Classification with Deep Convolutional Neural Networks. In F. Pereira, C. J. C. Burges, L. Bottou and K. Q. Weinberger (ed.), Advances in Neural Information Processing Systems 25 (pp. 1097--1105) . Curran Associates, Inc. 

[10] Saha S, Bhattacharjee A, Fattah SA. Automatic detection of sleep apnea events based on inter-band energy ratio obtained from multi-band EEG signal. Healthc Technol Lett. 2019;6(3):82-86. Published 2019 Jun 3. doi:10.1049/htl.2018.5101.

[11] J. Deng, W. Dong, R. Socher, L. Li, Kai Li and Li Fei-Fei, "ImageNet: A large-scale hierarchical image database," 2009 IEEE Conference on Computer Vision and Pattern Recognition, 2009, pp. 248-255, doi: 10.1109/CVPR.2009.5206848.

[12] Ferri R, Rundo F, Silvani A, Zucconi M, Bruni O, Ferini-Strambi L, Plazzi G, Manconi M. REM Sleep EEG Instability in REM Sleep Behavior Disorder and Clonazepam Effects. Sleep. 2017 Aug 1;40(8). doi: 10.1093/sleep/zsx080. PMID: 28482056.

[13] Schensted, C. (1961). Longest Increasing and Decreasing Subsequences. Canadian Journal of Mathematics, 13, 179-191. doi:10.4153/CJM-1961-015-3.

[14] Sasai-Sakuma T, Inoue Y. Differences in electroencephalographic findings among categories of narcolepsy-spectrum disorders. Sleep Med. 2015 Aug;16(8):999-1005. doi: 10.1016/j.sleep.2015.01.022. Epub 2015 Apr 1. PMID: 26026626.

[15] Siddiqui MM, Srivastava G, Saeed SH. Diagnosis of insomnia sleep disorder using short time frequency analysis of PSD approach applied on EEG signal using channel ROC-LOC. Sleep Sci. 2016;9(3):186-191. doi:10.1016/j.slsci.2016.07.002
[16] Zhao, Li and He, Yang. (2013). Power Spectrum Estimation of the Welch Method Based on Imagery EEG. Applied Mechanics and Materials. 278-280. 1260-1264. 10.4028/www.scientific.net/AMM.278-280.1260. 

[17] https://www.cdc.gov/nchs/nvss/index.html[Accessed 4.5.2021]

\end{document}